\documentclass{aa}
\usepackage{graphicx}

\begin{document}
   \title{Multi-periodic oscillations of $\alpha$ Hya
   \thanks{Based on observations at 2.2m-MPG/ESO in March--April 2005 at the ESO-La Silla observatory.}
   }
   \author{J. Setiawan\inst{1}
          \and
          P. Weise\inst{1,2}
          \and
          M. Roth\inst{3}
          }
   \offprints{setiawan@mpia-hd.mpg.de}

\institute{Max-Planck-Institut f\"ur Astronomie, K\"onigstuhl 17, 69117 Heidelberg, Germany 
	   \and Department of Physics and Astronomy, University of Heidelberg, Germany  
         \and Kiepenheuer-Institut f\"ur Sonnenphysik, Sch\"oneckstr. 6, 
79104, Freiburg (Brsg), Germany
}
	    
\date{Received 25 April 2005 / Accepted}
\abstract{We report the detection of multi-periodic oscillations of the cool 
evolved star $\alpha$~Hya (HD~81797, K3II-III). 
Two-hundred and forty-three high-resolution spectra ($R=48\,000$) 
of this star have been obtained in March and April 2005 with FEROS at 
the 2.2~m-MPG/ESO telescope in La Silla Observatory, Chile. 
We observed oscillations in the stellar radial velocity and the asymmetry 
of the spectral line profile. 
We detected oscillation frequencies of the stellar radial velocity 
in two frequency regions, $\nu= 2-30\, \mu$Hz and $\nu= 50-120\, \mu$Hz. 
The corresponding periods are $P= 0.6-5.6$ days and $P= 2.3-5.5$ hours, respectively. 
In addition to these oscillations we also observed 
a trend in the radial velocity which shows evidence for a long-term variability. 
Furthermore, our measurements show a correlation between the variation in the 
radial velocity and the asymmetry of the spectral line profile, 
as measured in the bisector velocity spans. 
The line bisectors also show oscillations in the same frequency regions as those of 
the radial velocity. 
We identified 13 oscillation frequencies in the bisector variation with 
equidistant separations of $8.09\pm0.32\,\mu$Hz in the lower frequency 
region and $7.69\pm0.47\,\mu$Hz in the higher frequency region.
The source of the short-term oscillations of $\alpha$~Hya is 
obviously due to non-radial stellar pulsations. 
These oscillations may have a similar origin like oscillations in solar-like stars.
The detection of the multi-periodic oscillations in $\alpha$-Hya makes 
this star to be an amenable target for asteroseismology, in particular, as it is a star 
in the red giant branch.
}
\maketitle

\section{Introduction}
Since the last two decades the precise radial velocity (RV) technique 
has been very successful in the detection of planets around stars other than the Sun. 
In fact, this technique gives also a possibility to study the 
stellar interiors by using seismology methods similar to 
helioseismology, which has been applied to probe the interior of 
the Sun by measuring low-amplitude solar oscillations. 

Stars with outer convection zones are in principle subject to ``solar-like'' oscillations 
because the turbulent motions in the outer convection zones could drive solar-like oscillations. 
The amplitudes of those oscillations are smaller compared to the pulsations of 
the Mira variable stars and are a challenge to detect. 
With the precise RV techniques several reports on the detection of solar-like 
oscillations have been made from ground (Bedding~\&~Kjeldsen~\cite{bed03} and references herein). 

In the 1980s red giant stars have been found to be variable stars by measuring the 
radial velocity (Walker~et~al.~\cite{wal89}; Smith~et~al.~\cite{smi87}). 
The amplitude of the variation in RV is between $30-300\,\mathrm{m\,s}^{-1}$. 
Further RV surveys showed that G and K giants exhibit RV variations on time scales 
from several hours up to hundreds of days. 
Multi-periodicities in the RV variations have also been reported 
(Hatzes~\&~Cochran~\cite{hat94a},b).  

The observed RV variability in red giant stars can either caused by the presence of 
stellar or sub-stellar companions, surface inhomogeneities, e.g., starspots, or 
stellar pulsations. 

Evidences for sub-stellar companions have been reported by several 
authors (see e.g., Hatzes~et~al.~2005 and references herein). These detections are 
complimentary to the detection of planetary companions around solar-like stars.

Concerning the rotational modulation in red giants, some methods have been 
used to measure stellar activity, e.g. by using chromospheric activity indicators 
such as Ca II H and K emission lines or the variation in spectral line bisector. 
Choi~et~al.~(\cite{cho95}) reported rotational modulation in a sample of K giants 
by analysing Ca II H and K emission lines. 
Setiawan~et~al.~(\cite{set04}) detected rotational modulation 
in the giant HD~78647 by examining the spectral line bisector.

Evidences for non-radial stellar oscillations in G and K giants have been 
detected, e.g., in Arcturus (Smith~et~al.~\cite{smi87}; Hatzes~\&~Cochran~\cite{hat94a}), 
$\beta$ Oph (Hatzes~\&~Cochran~\cite{hat94b}), 
$\alpha$ UMa (Buzasi~et~al.~\cite{buz00}) and $\xi$~Hya (Frandsen~et~al.~\cite{fra02}).\\
Dziembowski~et~al.~(\cite{dzi01}) presented a theoretical 
analysis of oscillations observed in red giants. 
According to stellar model calculations red giant stars can be subject to 
gravity and acoustic modes ($g$ and $p$ modes), however these models 
are not in perfect agreement with the observations. 
Especially, the driving mechanisms are still unclear. 
Besides the convective driving, Mira-like mechanisms are also possible with 
an decrease in oscillation amplitude with increasing frequency.
Further observations of stellar oscillations would be a valuable source for such asteroseismic 
investigations of stellar properties.

In this paper we report the detection of multi-periodic oscillations in 
$\alpha$~Hya. We observed short-term oscillations with periods of 
several days up to hours (Sect.~4). 
We also confirmed the presence of a long-period RV variation reported 
already earlier. However, based on our data we are still not able to give the 
best interpretation of this long-term variability.
In Sect.~5 we present our analysis 
of the spectral line profile asymmetry (bisector). We found a correlation 
between the variation in the bisectors and the RVs. 
This finding leads to the source of the observed short-period RV oscillations (Sects.~5,~6). 

\section{The star $\alpha$~Hya}
$\alpha$~Hya (HD~81797) is a cool evolved star. In the Hertzsprung-Russell 
diagram this star is located in the upper part of the red giant branch (RGB). 
The basic information and stellar parameters are listed in Table~1. 
\begin{table}[b]
\caption{Stellar parameters of $\alpha$~Hya.} 
\begin{tabular}{lll}
\hline
\hline
Identifier       & $\alpha$~Hya, HD~81797	   & {\sc Simbad}   \\
Spectral type    & K3II-III                        & {\sc Hipparcos}\\
Parallax         & $18.4\pm0.78$                   & mas \\
$m_v$            & 1.99                            & mag \\
$B-V$            & 1.44                            & mag \\
$BC$             & -1.002                          & mag \\
$T_\mathrm{eff}$ & 4086                            & K   \\
Angular diameter & $10.5\pm0.5$                    & mas \\
Radius           & $61.2\pm0.4$                    & R$_{\sun}$ \\
$v\sin{i}$       & $<$1.4                          & km/s \\
$[\mathrm{Fe/H}]$& -0.12 .. -0.19                  & dex \\
$\log{g}$        & 1.77 .. 1.86                    & cm/s$^2$ \\
Stellar mass     & $2.5-3$                         & M$_{\sun}$ \\
\hline
\hline
\end{tabular}
\end{table}

The spectral type, parallax, visual magnitude $m_v$ and 
color index $B-V$ were taken from the {\sc Hipparcos} and Tycho Catalogues, ESA (1997).
By using the {\sc  Hipparcos} photometry data and the calibration 
for effective temperatures $T_\mathrm{eff}$ and bolometric corrections $BC$ 
by Flower~(\cite{flo96}),  we computed a stellar radius of \mbox{$R \approx  61$~R$_{\sun}$ (0.29~AU)}.
We determined an angular diameter of $\Theta=$~10.5~mas, which is in good agreement 
with the values given in the CHARM and CHARM2 catalogues 
(Richichi~\&~Percheron~\cite{ric02}; Richichi~et~al.~\cite{ric05}).
The projected rotational velocity $v\sin{i}$ was computed with 
the cross-correlation technique (see Setiawan et al.~\cite{set04} for details). 

We adopted the metallicity [Fe/H] and the surface gravity $\log{g}$ 
from Cayrel~de~Strobel~(\cite{cay01}). 
The accurate mass determination of red giants is more difficult than 
that of main-sequence stars. Based on the evolutionary track by Girardi~et~al.~(\cite{gir00}) 
we estimated a stellar mass of $2.5-3$~M$_{\sun}$.

The RV variation of $\alpha$~Hya has been reported by Walker~et~al.~(\cite{wal89}), 
Murdoch~\&~Hearnshaw~(\cite{mur93}), Skuljan~et~al.~(\cite{sku00}) and Setiawan~et~al.~(\cite{set04}).
They found long-period RV variation with a period of several hundreds of days.
However, there is still no clear explanation about the nature of 
this long-term variability. 
It can be the result of rotational modulation due to surface inhomogeneities (starspots) or 
the presence of stellar/sub-stellar companion(s). 
The previous RV surveys of $\alpha$~Hya did not report evidence for short-period RV oscillations. 
This may be caused by the poor sampling of the data and/or the instrumental 
limitation in the velocity accuracy. 


\section{Observations and data reduction}

We observed $\alpha$~Hya with FEROS at the 2.2~m MPG/ESO telescope 
in La Silla observatory, Chile. 
The spectrograph resolution of FEROS is $R=$~48\,000. 
It has a wavelength coverage from $3500-9200$~\AA~ (Kaufer~\&~Pasquini~\cite{kau98}). 
FEROS is equipped with two fibres. 
In the science exposure, the first fibre is used to take the spectrum 
of the object (star), whereas the second fibre can be used either 
to take the sky background, or the calibration spectrum (ThAr+Ne) 
in the ``simultaneous calibration'' mode. 
The short-term (few weeks) velocity precision of FEROS is 5~$\mathrm{m\,s}^{-1}$.  

During eight consecutive nights from $17-24$ March 2005 (data set I)
and other four consecutive nights from 30 March $2005-2$ April 2005 (data set II) 
we acquired 243 high-resolution spectra of $\alpha$~Hya 
with the simultaneous calibration mode.
Since $\alpha$~Hya is a bright star \mbox{($m_{V} \approx 2.0$)}, under good observing condition, 
an exposure time of $20-30$~s was optimal to obtain a 
stellar spectrum with a signal-to-noise ratio of $\sim500-600$ in the wavelength region 
$\lambda= 5000-5500$~\AA~. 

\begin{figure}[t]
   \centering
   \includegraphics[width=7.5cm, height=6.5cm]{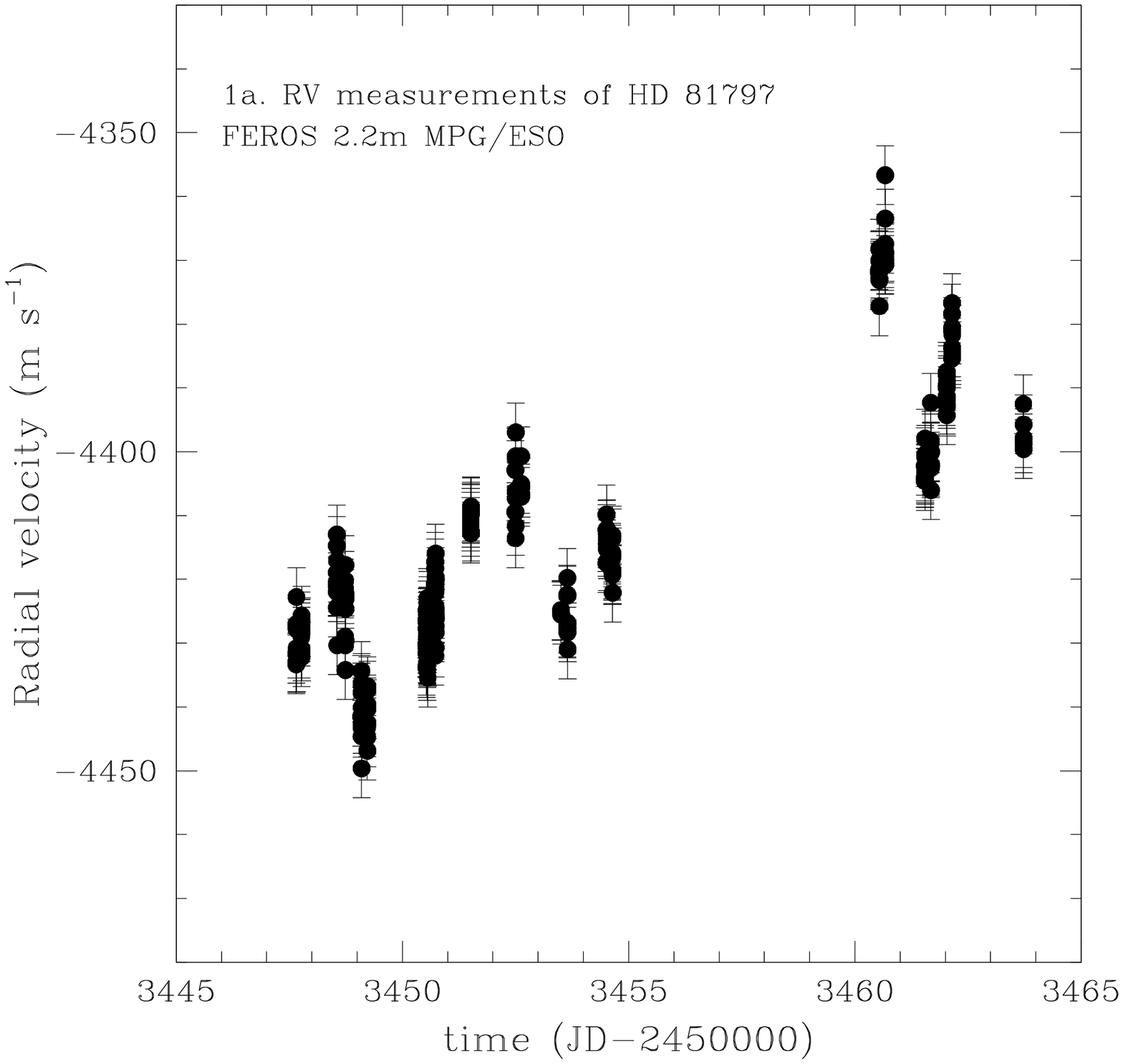}
   \includegraphics[width=7.5cm, height=6.5cm]{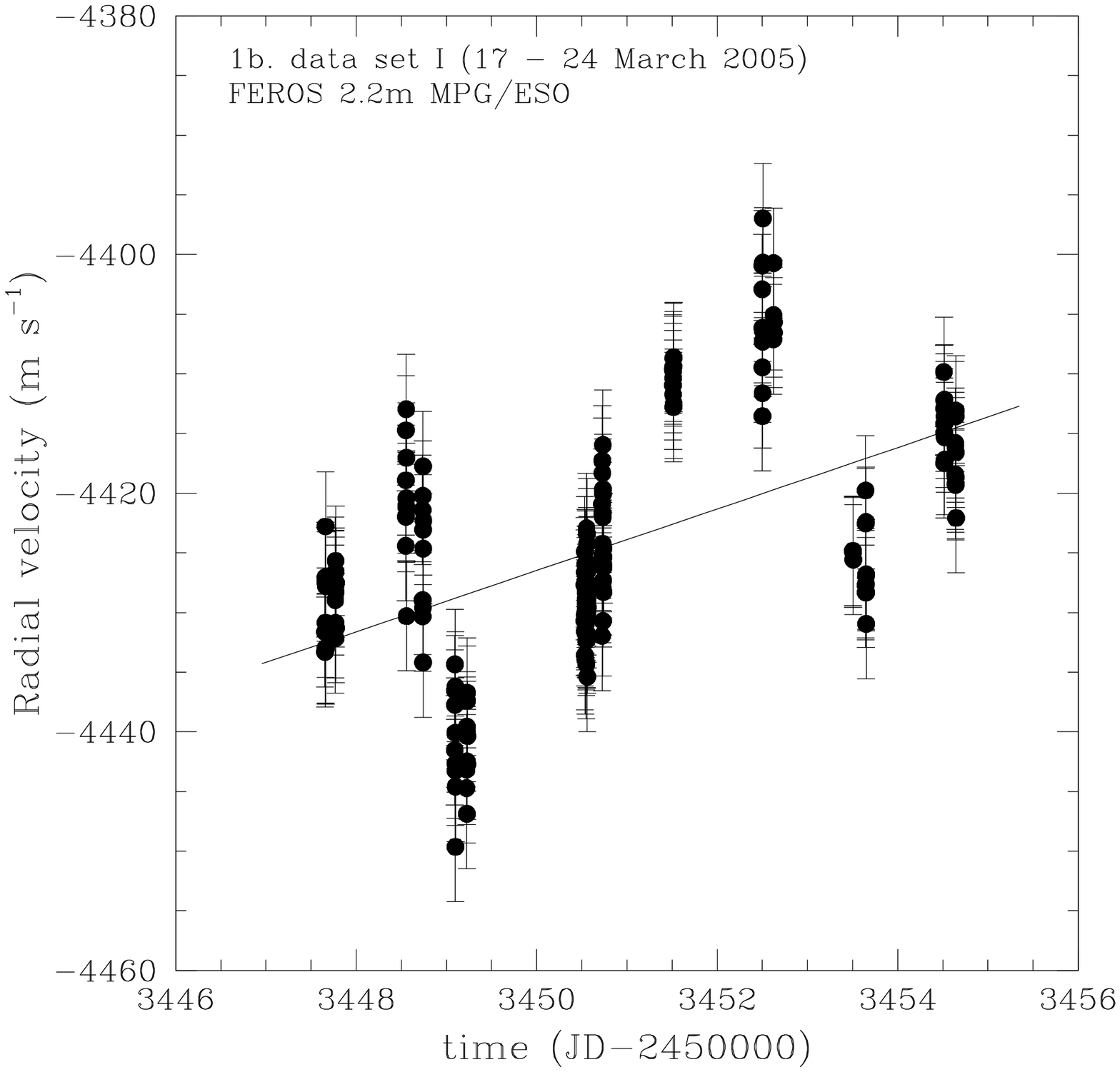}
   \includegraphics[width=7.5cm, height=6.5cm]{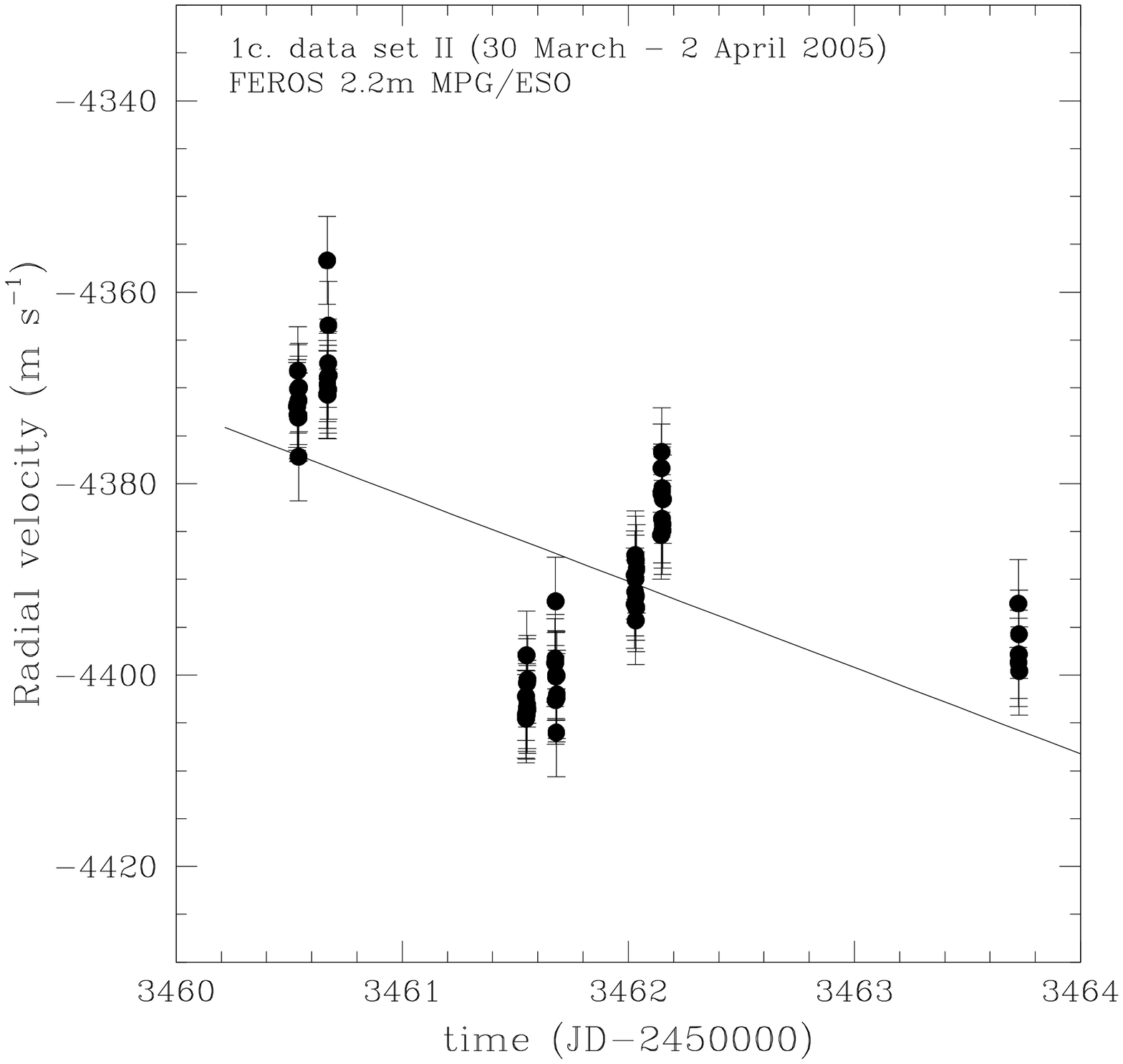}
   \caption{Radial velocity variation of $\alpha$~Hya. 
   To avoid effects on the oscillation frequencies due to the observational gap, 
   the whole RV measurements (a) will be treated separately in data set I (b) and data set II (c).}             
   \label{fig1}
\end{figure}

In order to resolve the oscillations with time scale of minutes and hours, 
we recorded several time series of exposures. 
The time needed for a series of 10 spectra including the CCD-readout time 
was $10-12$ minutes ($\approx$50 minutes for a series of 40 spectra). 
The time interval between the series was $1-6$ hours.

The data reduction has been carried out by using the FEROS-DRS pipeline, 
which produced 39 one-dimensional wavelength calibrated spectra for each fibre.  

\section{Radial velocity variation}
We measured the RVs by using the cross-correlation technique. 
The stellar spectra were cross-correlated with a numerical template (Baranne~et~al.~\cite{bar96}). 
The cross-correlation function was then fitted with a Gaussian.
The RV was obtained from the position of the dip of the cross-correlation function.
The detailed procedures of the computation were described in Setiawan~et~al.~(\cite{set03}). 

Fig.~1 shows the RV measurements of $\alpha$~Hya. As seen in the Fig.~1a 
there is an observational gap between the two data sets. 
The RVs show variations with time scale of days. 
By treating the data set I (Fig.~1b) and II (Fig.~1c) separately, 
a linear trend can be recognized by visual inspection in each data set.

We also observed a long-term RV variation. 
Our measurements did not cover the full period of this long-term variability. 
Therefore, we are not able to find the best model which may 
fit the whole long-period RV oscillation. 
Such a fit-function, when subtracted from the RVs, would allow to compute the residual velocities.  
Nevertheless, it is possible to treat the data sets I and II separately 
and subtract the respective linear trends from the two sets of RV measurements. 
The then calculated residual velocities of each data set are displayed in Fig.~2a,b.
The figures show oscillations of the residual with a maximal velocity amplitude 
of $\sim$20~$\mathrm{m\,s}^{-1}$.  

We computed a Lomb-Scargle (LS) periodogram (Lomb~\cite{lom76}; Scargle~\cite{sca82}) 
of each data set to find the oscillation frequencies. 
Fig.~3a and 3b show the respective LS periodograms of the residual RV measurements. 
We identified significant oscillation frequencies in two regions, 
i.e., $\nu= 2-30\,\mu$Hz and $\nu= 50-120\,\mu$Hz in both data set I and also in the 
data set II. 

\begin{figure}[t]
  \centering
  \includegraphics[width=7.5cm, height=6.5cm]{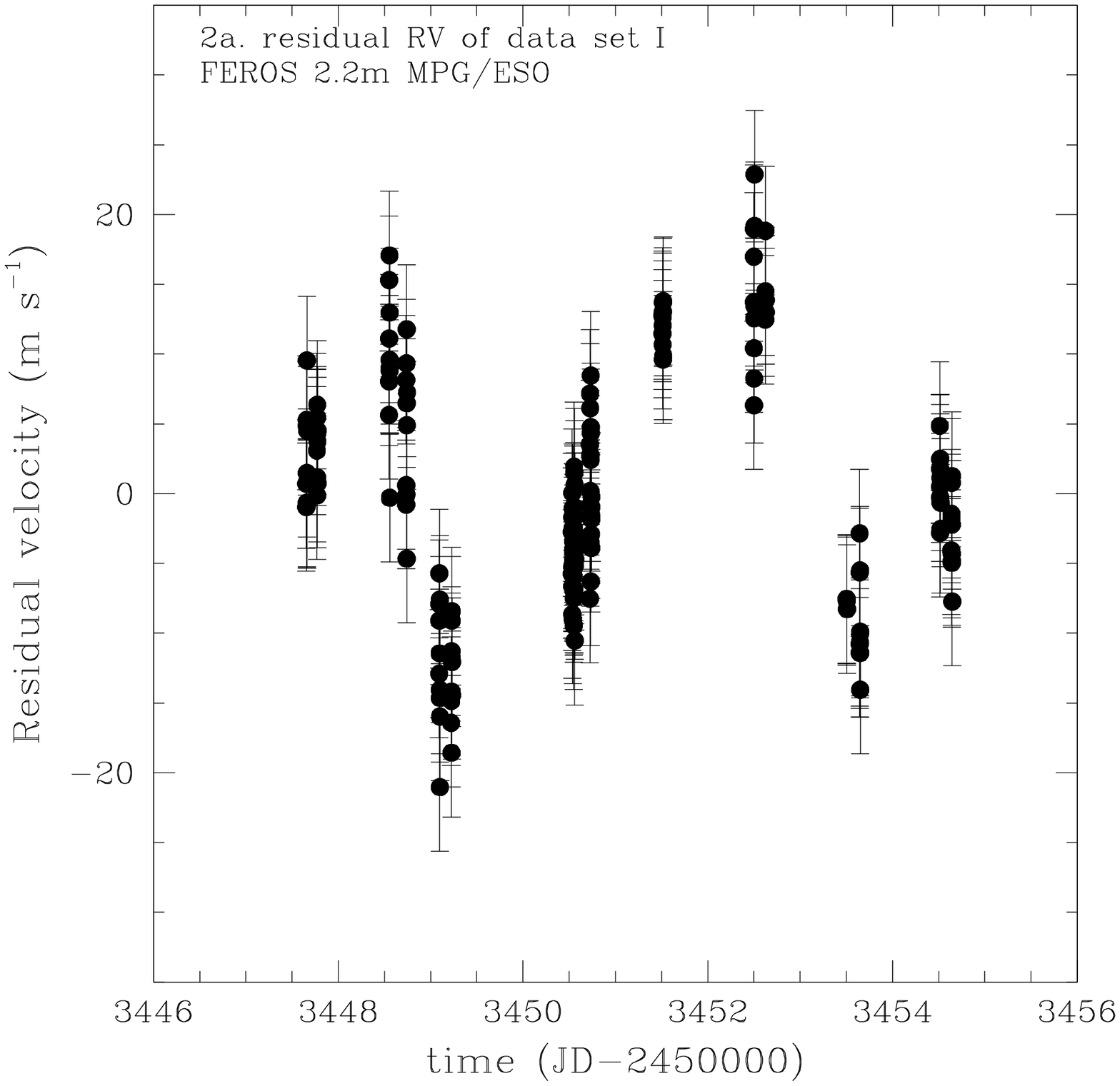}
  \includegraphics[width=7.5cm, height=6.5cm]{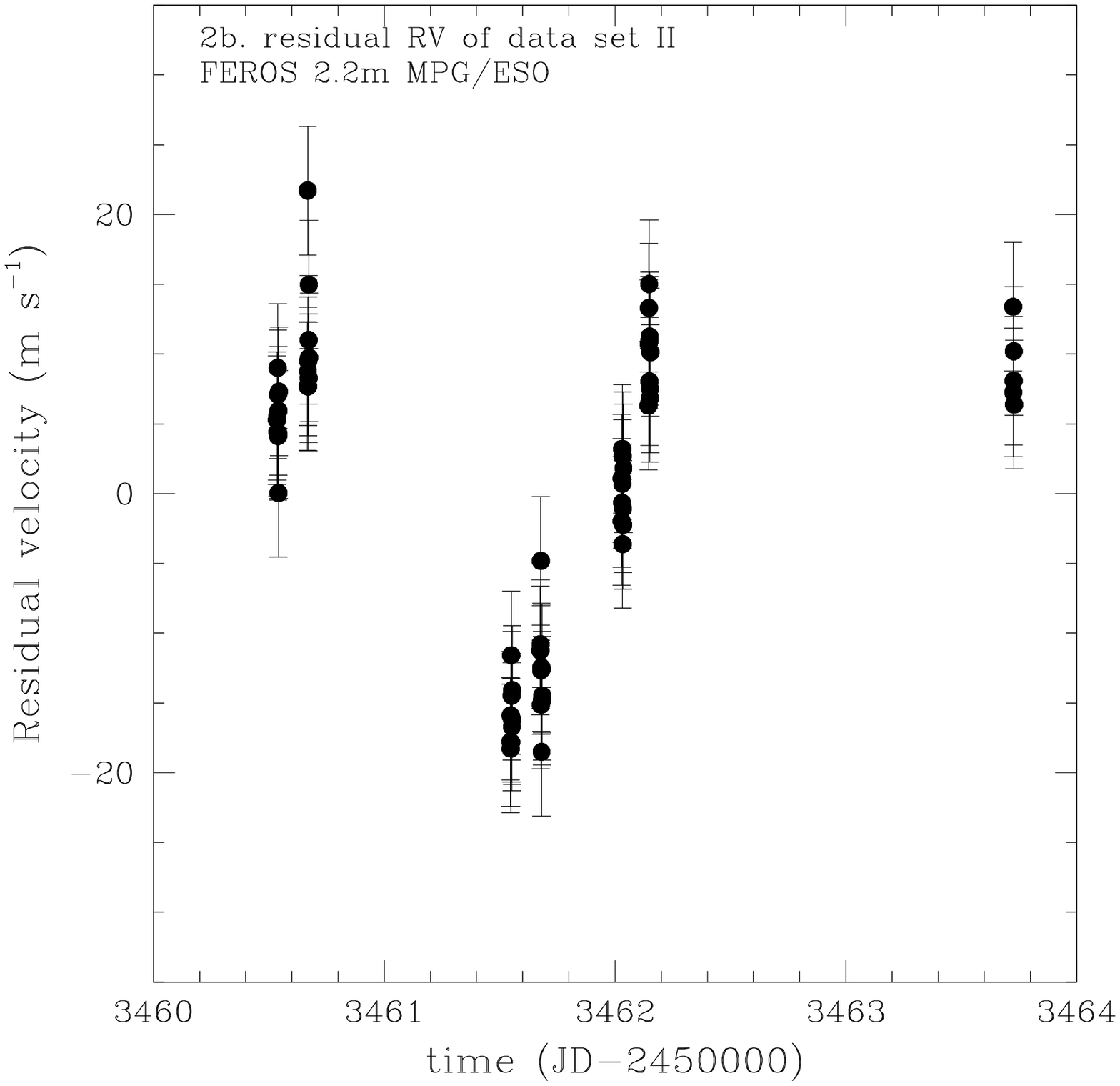}
  \caption{Residual RV of $\alpha$~Hya for the data sets I (upper panel) and II (lower panel).}             
   \label{fig2}
\end{figure}

\begin{figure}[t]
   \centering
   \includegraphics[width=7.5cm, height=6.5cm]{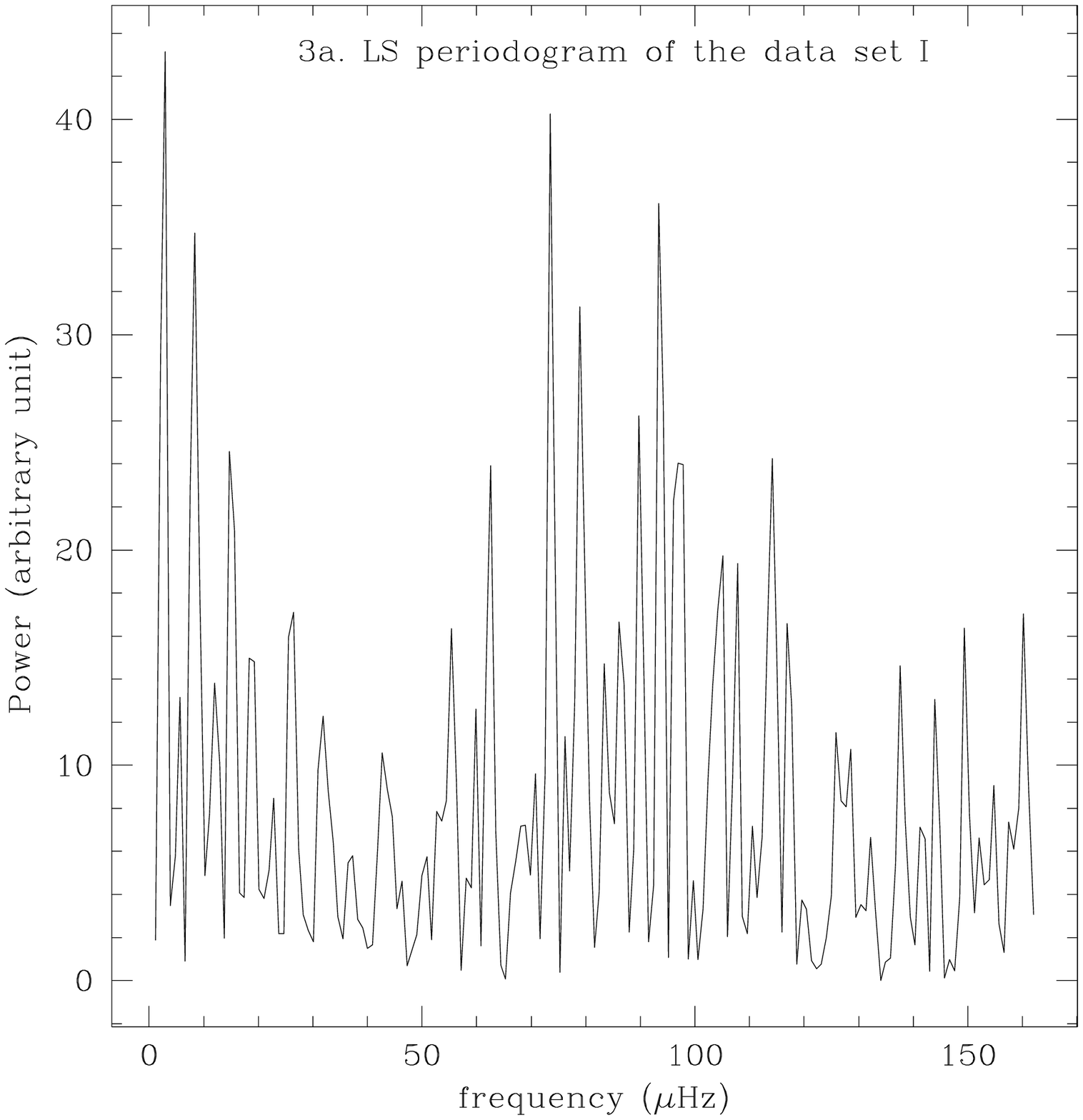}
   \includegraphics[width=7.5cm, height=6.5cm]{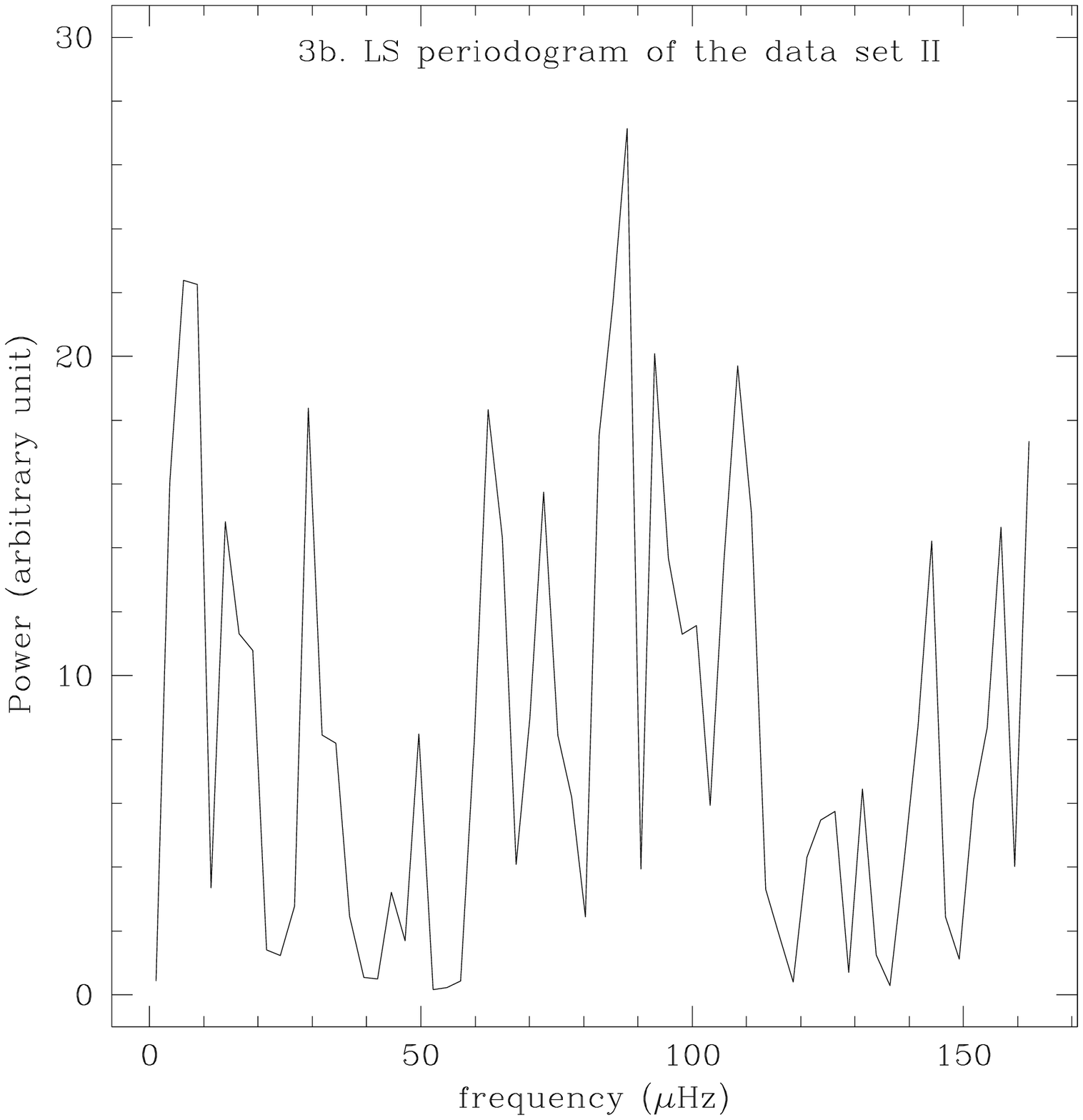}
   \caption{Lomb-Scargle periodogram of the residual velocities of $\alpha$~Hya for 
   data set I (upper panel) and data set II (lower panel)}
   \label{fig3}
\end{figure}

In Table~2 we list the frequencies and the corresponding periods 
obtained from the periodogram of the residual velocities. 
The LS periodogram of data set I shows the highest power 
at $\nu= 2.513\,\mu$Hz ($P=$ 4.61 days). 
Due to the observational time coverage of data set II the peak at this low frequency however, 
could not be observed, but it is merged in the peak at $\nu=7.530\,\mu$Hz.
In both data sets we found periods around 1.5 days, 4.4 hours and 3.0 hours. 
These periods are also found in the line bisector variation (see Sect.~5). 
The oscillation frequencies in the region $\nu= 50-120\,\mu$Hz are not equidistant. 
A possible explanation is, that the ``missing'' oscillation frequencies still 
have not been detected yet. 
This could have been occurred if the RV variation amplitudes of these modes were too small, i.e., 
beyond the accuracy of the spectrograph. Longer observations will be necessary to identify these oscillations, too.

\begin{table}[b]
\caption{Identification of oscillation frequencies of $\alpha$~Hya 
from the periodogram of the residual velocities} 
\begin{tabular}[h]{rrlrl}
\hline
\hline 
     & {\bf Data set I}&             & {\bf Data set II}  &            \\
$N$  & $\nu$ [$\mu$Hz] &  $P$ [day]  & $\nu$ [$\mu$Hz]    & $P$ [day]   \\
\hline
1    & 2.51           &  4.61       & -                  &             \\
2    & 8.07           &  1.43       & 7.53               & 1.54        \\
3    & 14.72          &  0.79       & -                  &             \\
4    & -              &  -          & 29.25              & 0.40        \\
\hline
$N$  & $\nu$ [$\mu$Hz] &  $P$ [hours]& $\nu$ [$\mu$Hz]   & $P$ [hours] \\
\hline
5    & 62.62          &  4.44       & 62.45              & 4.45         \\
6    & 73.46          &  3.78       & -                  & -            \\
7    & 78.89          &  3.52       & -                  & -            \\
8    & 89.73          &  3.10       & 86.71              & 3.20         \\
9    & 93.80          &  2.96       & 93.09              & 2.98         \\
10   & 96.96          &  2.87       & -                  & -            \\
11   & -              &  -          & 108.41             & 2.56         \\
12   & 114.14         &  2.43       & -                  & -            \\
\hline
\hline
\end{tabular}
\end{table}
\section{Spectral line profile asymmetry}

Surface inhomogeneities such as starspots or/and granulation will 
leave imprints in the spectral lines. 
They introduce an asymmetry in the spectral line profile (Gray~\cite{gra82}). 
This feature can be measured from the shape of the line bisector. 
An equivalent method to measure the variation in the line profile's asymmetry 
is to measure the bisector velocity span (Hatzes~\cite{hat96}). 
Non-radial oscillations are expected to be also mapped into the bisector velocity span (BVS). 
If oscillations like $p$- and $g$-modes are present, then they would result in a temporal 
variation of the BVS. 

The measurement of the BVS can be done as well with the 
cross-correlation method, as demonstrated by Queloz~et~al.~(\cite{que01}).
We computed the bisector velocity spans of the cross-correlation profile. 
We observed a time variation of the BVS. 
When plotted the BVS against the RV, we found a correlation between them (Fig.~4). 

This correlation indicates that the observed RV variation 
is most likely due to intrinsic processes in the star, i.e., (non-radial) 
stellar pulsations.

Other possibilities, like companions and rotational modulation due to starspots can be excluded.  
Because, a companion with an orbital period of a time scale of few days would have an orbit 
which is smaller than the stellar radius of $\alpha$~Hya. 
A rotational modulation of the period of several days would imply, 
that the star should rotate very fast ($v\sin{i}>100$~$\mathrm{km\,s}^{-1}$). 
Such a fast rotating K giant would have a strong X-ray emission and thus, 
it should have been detected by the ROSAT survey.

\begin{figure}[t]
   \centering
   \includegraphics[width=7.5cm, height=6.5cm]{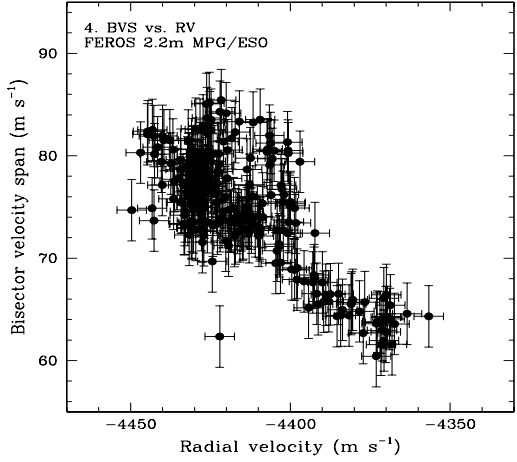}
      \caption{BVS vs. RV of $\alpha$~Hya. The figure shows a correlation between the BVS and RV.}
         \label{fig4}
   \end{figure}

The detection of oscillations in the time series of the BVS (Fig.~5a) is an 
observational approval of the method proposed by Hatzes (\cite{hat96}). 
There, it was demonstrated, that pulsations affect the shape of the spectral line 
profile and can be measured as variation in the bisector. Depending on the 
azimuthal order of a mode, the bisector is more or less affected.
\begin{figure}[h]
   \centering
   \includegraphics[width=7.5cm, height=6.5cm]{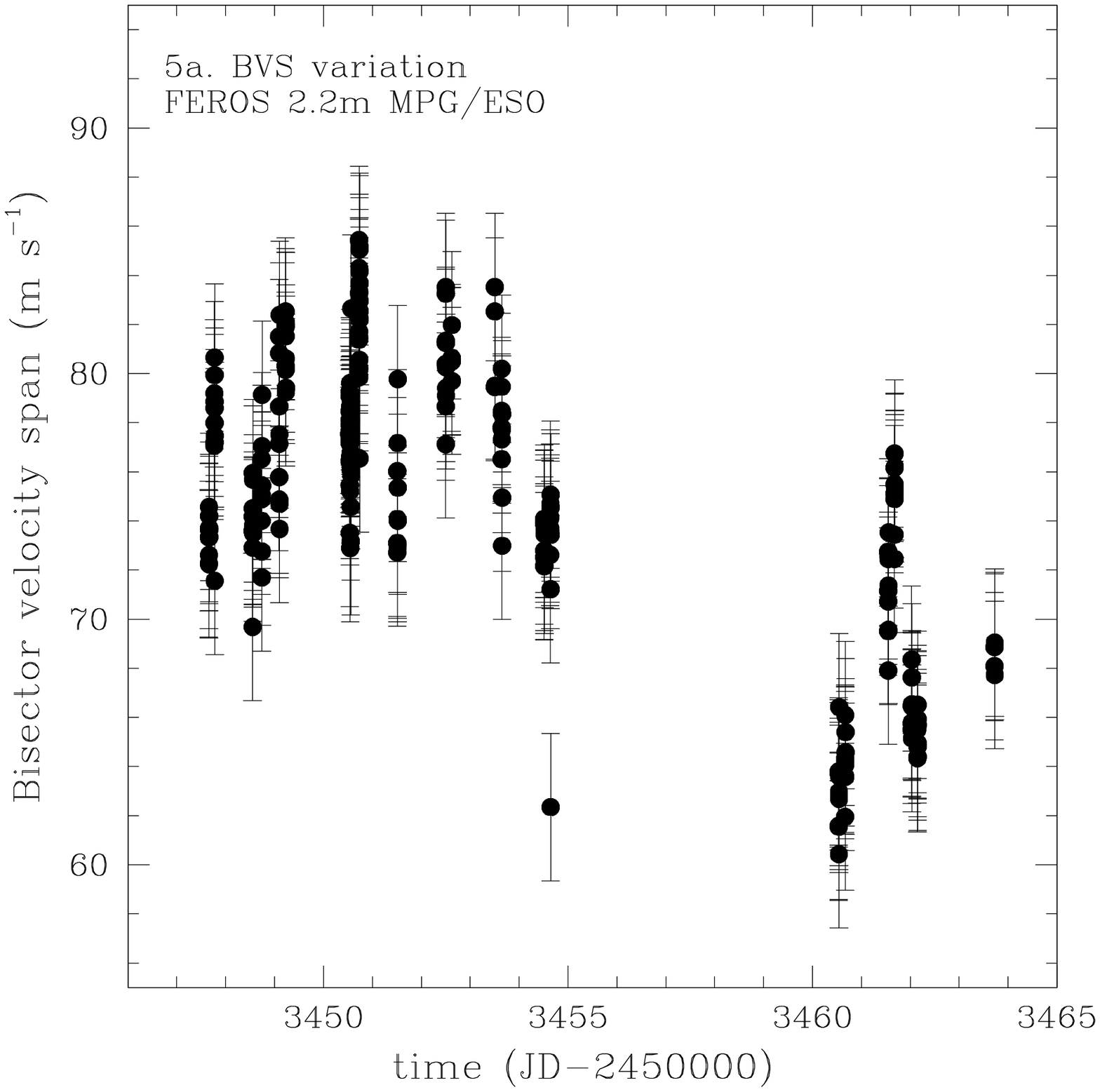}
   \includegraphics[width=7.5cm, height=6.5cm]{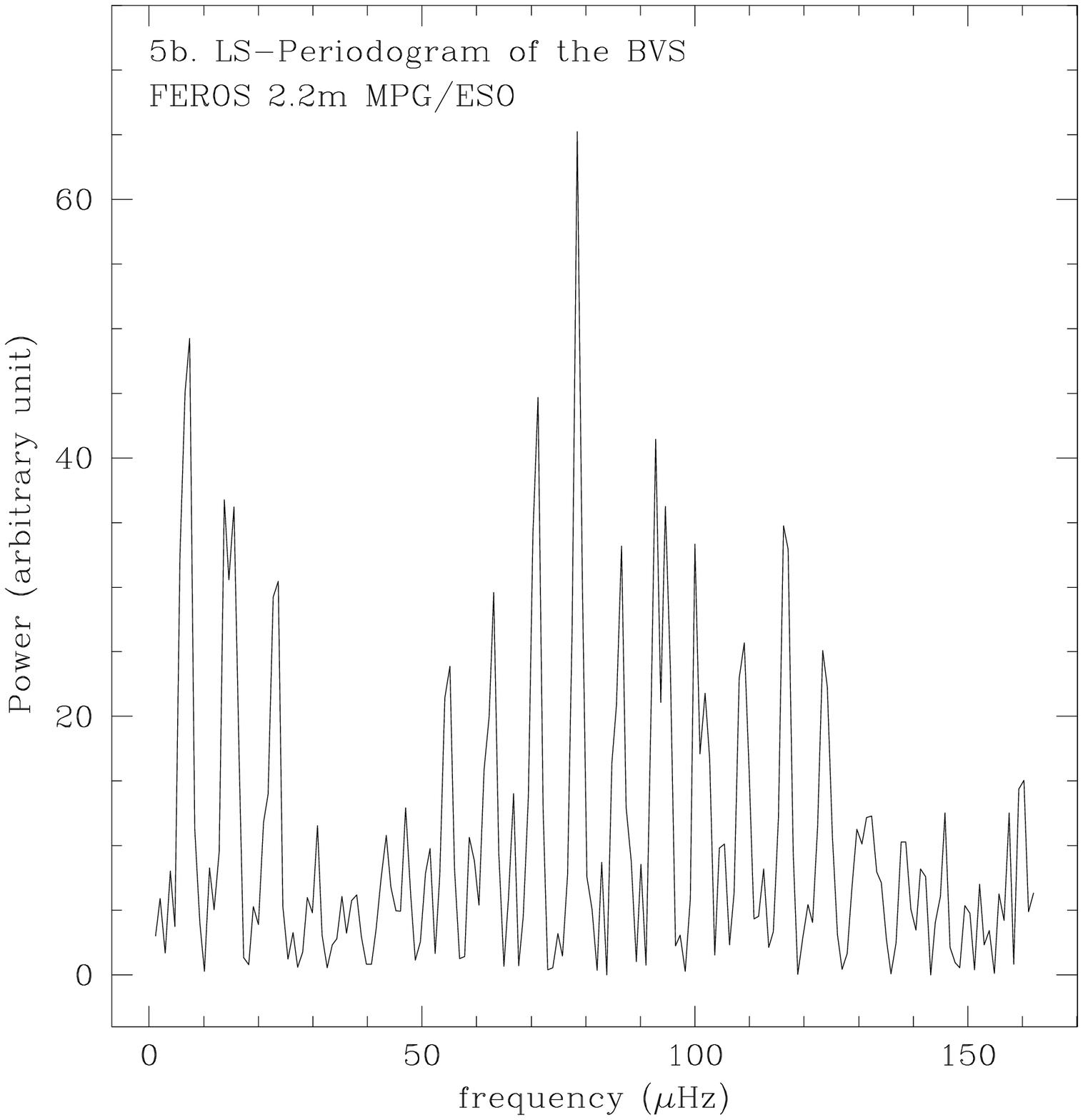}
     \caption{Upper panel: time variation of bisector velocity span of $\alpha$~Hya, 
     lower panel:Lomb-Scargle periodogram of the bisector velocity span. }
         \label{fig5}
   \end{figure}

\begin{table}[h]
\centering
\caption{Identification of oscillation frequencies of $\alpha$~Hya 
from the periodogram of the BVS}
\begin{tabular}{llll}
\hline
\hline 
$N$  & $\nu$ [$\mu$Hz] &  $P$ [days] \\
\hline
 1 & 6.550    &    1.77 &  \\
 2 & 14.639   &    0.79 &  \\
 3 & 23.177   &    0.50 &  \\
 \hline
$N$  & $\nu$ [$\mu$Hz] &  $P$ [hours] \\
 \hline
 4 & 54.63   &    5.08  &  \\
 5 & 63.17   &    4.40  &  \\
 6 & 70.81   &    3.92  &  \\
 7 & 78.45   &    3.54  &  \\
 8 & 86.54   &    3.21  &  \\
 9 & 93.73   &    2.96  &  \\
10 & 100.92  &    2.75  &  \\
11 & 108.56  &    2.56  &  \\
12 & 116.65  &    2.38  &  \\
13 & 123.84  &    2.24  &  \\
\hline
\hline
\end{tabular}
\end{table}
The LS periodogram of the whole determined BVS time series (data set I and data set II) 
shows peaks in the frequencies between $5-30\,\mu$Hz and $50-120\,\mu$Hz (Fig.~5b). 
As in the case of the RV measurements, the same frequency regions are covered in the 
BVS periodogram. The highest peaks are at 6.55~$\mu$Hz and 78.45~$\mu$Hz. 
The corresponding periods are $\approx$ 1.8 days and $\approx$ 3.5 hours, respectively.
Moreover, the peaks are equidistant in the smoothed periodogram. 
The frequency distance of the major peaks in the lower peak region is on 
average $8.09\pm0.32$~$\mu$Hz, whereas in the higher peak region the average peak 
distance is $7.69\pm0.47$~$\mu$Hz. The list of the identified peaks with 
a false alarm probability of FAP$\le 10^{-7}$ is given in Table~3. 

The oscillations in the spectral line asymmetry, and in particular 
the correlation between BVS and RV indicates, that the observed variations are 
due to non-radial oscillations. 
The shape of the power distribution in the periodograms of the RV and BVS 
measurements with high amplitudes at high frequencies indicates that there 
exists the possibility of convectively driven modes in the frequency range 
around 78.5~$\mu$Hz. 
These modes therefore are rather solar-like oscillations than 
Mira-like pulsations (Dziembowski~\cite{dzi01}). 
The modes around 10~$\mu$Hz in the RV and the BVS periodograms seem to exhibit 
a decrease in amplitude with higher frequencies, which would support 
a Mira-like interpretation for these pulsations. 
Nevertheless having the solar case in mind, it might also be possible, 
that these are $g$-modes. However, a final statement can only be given 
on the basis of longer data sets and on stellar models of $\alpha$~Hya. 

\section{Discussion and conclusion}
Only little is understood about the nature of short-period oscillations of evolved stars. 
Especially, the excitation of the modes itself and the interaction with 
convection might be different from what is known of solar oscillations. 
However, this fact is based on theoretical investigations. 

Until now, nonradial pulsations in red giants have been detected 
only in a few numbers of stars. 
Therefore, each detection will give a valuable contribution to asteroseismology, 
which is a powerful tool for performing direct tests of stellar structure and evolution theory. 
The identification of many pulsation frequencies in a star 
would provide a wealth of information, which can be obtained by 
relating the oscillation spectra to stellar physical properties.  
Moreover, from a lot of oscillation modes the interior of the star could be sounded.
Nevertheless, a lot of development is still necessary until it 
becomes possible to determine stellar parameters as accurately as the solar counterparts. 
One important step is the surveying of the stars for their 
pulsations and detecting amenable targets for asteroseismology.

We report of such a possible target, as we detected multi-periodic 
oscillations in $\alpha$~Hya.
The detection of these oscillations is twofold. 
On the one hand we found them in radial velocity measurements and on 
the other hand in measurements of the bisector velocity span.
Both measurements correlate well. 
Especially, this correlation hints to the detection of non-radial oscillation modes. 
Our observations are a direct approvement of 
the mapping of the oscillation velocities onto the line profile 
bisector as suggested by Hatzes~(\cite{hat96}). 

The next step will be an extension of RV and line sector asymmetry 
observations of $\alpha$ Hya in order to be able to perform a stability analysis of the modes.
As pointed out by Dziembowski et al.~(\cite{dzi01}) this would help 
to decide about the possible driving mechanism of the modes. 
Mira-like and solar-like processes are on debate.
A Mira-like interpretation would be related with an amplitude 
decrease towards higher frequencies.

Based on our data we find such a decrease only in a low frequency 
region around 10~$\mu$Hz. But the amplitudes raise again 
at higher frequencies around 78.5~$\mu$Hz in all periodograms 
calculated from the RV measurements and as well in the bisector 
velocity span measurements. 
Moreover the peaks are equidistant with a frequency difference 
of {\bf $8.09\pm 0.32$}~$\mu$Hz in the low frequency region 
and $7.69\pm 0.47$~$\mu$Hz in the higher frequency region. 
Therefore, these pulsations are likely to be solar-like oscillations. 

Our findings in $\alpha$~Hya are in good concordance with earlier 
detections of solar-like oscillations in red giant stars. 
Frandsen et al.~(\cite{fra02}) reported of radial oscillations 
in $\xi$~Hya in a frequency range between 50--130~$\mu$Hz. 
The detected peaks were also equidistant with a separation of 7.1~$\mu$Hz.
However, we also find evidences for non-radial and low-frequency 
oscillations in $\alpha$~Hya by analysing additionally the bisector velocity span.
Both stars $\alpha$ and $\xi$~Hya have masses close to 3~M$_{\sun}$. 
The progenitor main-sequence star of $\alpha$~Hya is probably an early A-type star 
(see Schmidt-Kaler~\cite{sch82}). 
Therefore, an agreement in the oscillation frequencies and 
in the separation of the peaks is very likely, as both stars 
might have underwent a similar evolution. 
Further extensive theoretical modelling of the finding of 
oscillations in $\alpha$~Hya is beyond the scope of this report 
and will be presented in a forthcoming paper.

\begin{acknowledgements}
J.~S thanks J. Pritchard for his works on FEROS. 
We kindly thank the 2.2m~MPG/ESO team for their assistances during 
the observations. J.~S thanks A.P. Hatzes and J. Rodmann 
for the discussions.
\end{acknowledgements}

\end{document}